\documentclass{article}
\usepackage[top=4cm, bottom=4cm, left=3cm, right=3cm]{geometry}
\usepackage{graphics}
\usepackage{epsfig}
\usepackage{enumerate}
\usepackage{amsmath, amssymb, amsfonts}
\usepackage{xcolor}
\usepackage[latin1]{inputenc}
\usepackage[T1]{fontenc}
\usepackage{pifont}
\usepackage{subcaption}
\usepackage{graphicx,animate}
\usepackage{hyperref}
\usepackage{comment}

\numberwithin{equation}{section}

\newtheorem{theo}{Theorem}[section]

\newtheorem{definition}{Definition}[section]
\newtheorem{rem}{Remark}[section]

\newenvironment{proof}[1][Proof]{\textbf{#1.} }{\ \rule{0.5em}{0.5em}}

\def \N {\mathbb{N}}
\def \P {\mathbb{P}}

\def \dis {\displaystyle}
\def \V {{\mathcal{V}}}

\def \M {{\mathcal{M}}}

\begin{document}
  \title{Stationary and Non-Stationary Transition Probabilities in Decision Making: Modeling COVID-19 Dynamics}
 \author{
 	{{Romario Gildas Foko Tiomela} \thanks{{\it
 				Department of Mathematics, EMERGE, NEERLab, Morgan State University, Baltimore, MD 21251, USA,\\
 				Email~: {\sf romario.foko@morgan.edu} }}}
        {{~~Samson Adekola Alagbe} \thanks{{\it
 				Department of Mathematics, EMERGE, Morgan State University, Baltimore, MD 21251, USA,\\
 				Email~: {\sf saala26@morgan.edu} }}}
 	{{~~Olawale Nasiru Lawal} \thanks{{\it
 				Department of Mathematics, EMERGE, Morgan State University, Baltimore, MD 21251, USA,\\
 				Email~: {\sf ollaw5@morgan.edu} }}}\\
                {{~~Serges Love Teutu Talla} \thanks{{\it
 				Department of Mathematics, EMERGE, Morgan State University, Baltimore, MD 21251, USA,\\
 				Email~: {\sf seteu1@morgan.edu} }}}
        {{~~Isabella Kemajou-Brown} \thanks{{\it
 				Department of Mathematics, EMERGE, Morgan State University, Baltimore, MD 21251, USA,\\
 				Email~: {\sf isabella.brown@morgan.edu} }}}
 }

  \date{\today}
  
  \maketitle	

\begin{abstract}
\noindent This study introduces a comparative modeling framework using stationary and non-stationary transition probabilities within a Markov Decision Process (MDP) to assess COVID-19 disease dynamics. Stationary transition probabilities assume constant transition rates, while non-stationary transitions reflect time-dependent behaviors including policy interventions or behavioral changes. We develop a comprehensive compartmental model with transitions based on binomial and multinomial processes. Mathematical models for both stationary and non-stationary transition frameworks are developed and simulated over a 365-day period to emphasize dynamic variations in epidemic outcomes. Our findings highlight the significance of non-stationary modeling in accurately representing the dynamic characteristics of pandemic situations and provide recommendations for optimizing public health interventions under uncertainty. This comparative analysis offers useful information for epidemiological modeling and decision-making in dynamic risk environments.
\end{abstract}


\section{Introduction}
\noindent Infectious disease modeling has a rich history, evolving from simple mathematical frameworks to sophisticated models incorporating various biological and social factors. To provide solid estimates, models need to be properly calibrated based on empirical evidence (see for instance \cite{biggerstaff2020early, chinazzi2020effect, crunfli2021sars}). One of the earliest models, the Reed-Frost model, employed a discrete-time approach to describe the probability of an individual being infected in successive time intervals. This model laid the foundation for subsequent studies that sought to capture the complexity of epidemic spread through probabilistic means. For some applications and generalization of the Reed-Frost model, see for instance \cite{barbour2004approximating, lefevre1990non, ng1990generalized,  perez2002simulation, ranta1999predicting, tsutsui2003stochastic}. Classical compartmental models such as SIR and SEIR generally assume constant transition rates between disease states \cite{anderson1991infectious, hethcote2000mathematics}. However, for pandemics such as Covid-19, these assumptions often break down due to dynamic changes such as: behavioral changes, fluctuation in healthcare capacities or public health policies \cite{biggerstaff2020early, giordano2020modelling, purkayastha2021comparison, tuckwell2007some}. Consequently, incorporating non-stationary transition probabilities enables a more realistic epidemic dynamics.
\\

\noindent
In \cite{palopoli2023markovian}, L. Palopoli et al. proposed a Markovian stochastic approach to model the spread of a SARS-CoV-2-like infection within a closed human group using a Partially Observable Markov Decision Process (POMDP). The primary objective was to model in detail the effects of transitions between different states and resource limitations, such as hospital beds and the daily availability of tests. Such models are primarily control-oriented, emphasizing the impact of decision commands on transition probabilities and the epidemic's progression.\\

\noindent
When it comes to applications, one notable use is the ability to compute exact probabilities and properties of interest given a control policy, such as the decision to implement a lockdown based on the estimated number of infected individuals. This includes determining the probability that the number of deceased individuals will exceed an acceptable threshold or evaluating more general properties expressed in propositional temporal logic. Additionally, it is possible to synthesize control policies that inherently respect these properties. For more literature on these applications, we refer to \cite{ahmadi2020control, anderson1991infectious,browne1988characterizing, giordano2020modelling, miller2012edge} and the references therein.\\

\noindent
In 2006, H. Tuckwell and R. Williams \cite{tuckwell2007some} investigated the properties of a simple discrete-time stochastic epidemic Markovian SIR model, in which the total population remained constant and individuals met a random number of others at each time step. In their model, individuals remained infectious for some time units, after which they became either removed or immune. The transition probabilities from susceptible to infected states were determined using the binomial distribution. Their findings have practical applications in controlling the size and duration of epidemics, thereby reducing their human and economic costs. This model offers a more realistic characterization of epidemics compared to classical discrete-time models, such as the Reed-Frost model, which is often used for analyzing agricultural epidemics.\\

\noindent In \cite{llopis2023estimating}, the authors used multi-state models to estimate transition probabilities between different health states in elderly patients, providing insights into the temporal progression of diseases and the factors influencing these transitions. Complementing this, A. J. Black \cite{black2015computation} discussed methodologies for determining the final size of an epidemic using stochastic processes, emphasizing the importance of accurately modeling the pathways to the absorbing state, which signifies the epidemic's end. Similarly, C. Barril et al. \cite{barril2023final} explored the final infection size in models that consider asymptomatic transmission, highlighting the implications of transition probabilities between states for understanding the severity and spread of an epidemic. Additionally, S. Purkayastha et al. \cite{purkayastha2021comparison} compared various epidemiological models in their ability to predict the growth rate and final size of the COVID-19 pandemic in India, focusing on how different models handle transition probabilities and the associated uncertainties in projections. For more information on models that have proven useful in determining the major factors affecting the growth rate and final size of an epidemic, see \cite{giles1977mathematical, hethcote2000mathematics}.\\

\noindent
Recently in 2021, A. Zardini et al. \cite{zardini2021quantitative} employed statistical methods to quantify the probability of transition between different states of COVID-19-affected patients based on age class. The authors provided estimates of the probabilities of transition across the stages characterizing clinical progression after SARS-CoV-2 infection, stratified by age and sex, as well as the time delays between key events. They analyzed a sample of 1,965 SARS-CoV-2 positive individuals who were contacts of confirmed cases. These individuals were identified irrespective of their symptoms as part of contact tracing activities conducted in Lombardy, Italy, from March 10 to April 27, 2020. They were also monitored daily for symptoms for at least two weeks after exposure to a COVID-19 case and tested for SARS-CoV-2 via real-time PCR. Additionally, F. Riccardo et al. \cite{riccardo2020epidemiological} evaluated the effects of lockdown policies in Italy following the outbreak of the pandemic.\\

\noindent
The current study addresses the critical need for robust models in infectious disease research, particularly focusing on COVID-19. By examining both stationary and non-stationary transition probabilities, this research provides a detailed understanding of disease dynamics over time. On the one hand, stationary transition probability
$$\P_{i,j}=\P(X_{q+\triangle q}=j|X_q=i)$$ 
represents the probability of transitioning from state $i$ to state $j$ within a fixed time period, where $X_q$ is the state at stage $q$ and $\triangle q$ the stage difference. This formulation is essential for understanding how diseases progress under constant conditions. On the other hand, non-stationary transition probabilities 
$$\P_{i,j}(t)=\P(X_{q+\triangle q}=j|X_q=i,~t)$$
allow for changes in transition probabilities over time, reflecting more realistic scenarios where the probability varies with respect to $t$ (with $t$ representing different intervention measures and other factors). This approach is crucial for modeling dynamic public health responses and understanding how interventions can alter the course of an epidemic.\\

\noindent
Stationary models serve as a useful baseline for evaluating the effects of new interventions by providing a consistent reference point. However, these models may misestimate impacts if the effects of interventions or the underlying disease dynamics change over time. Policymakers can use stationary models to identify deviations from expected outcomes, which may signal a need for policy adjustments. In contrast, non-stationary models offer a more dynamic approach by incorporating time-varying transition probabilities. This adaptability allows for real-time adjustments based on the latest data and evolving conditions. By reflecting changes in disease patterns, intervention effectiveness, and external factors, we anticipate that non-stationary models will enable more responsive and accurate decision-making. Our primary goal is to assess how different assumptions about transition dynamics impact model outcomes and their relevance to public health policy.\\

\noindent
The main contributions of this study include developing an epidemiological model with seven epidemiological compartments reflecting key clinical stages of COVID-19. We then derive probabilistic transition rules between states using binomial and multinomial distributions, and explicitly distinguish between stationary and non-stationary regimes. We define time-dependent transition parameters using logistic and quadratic forms to reflect real-world dynamics such as behavioral adaptation or resource saturation. Finally, we simulate COVID-19 dynamics under both transition assumptions and compare outcomes over a 365-day period, illustrating the impact of model choice on disease spread over time. We offer policy-relevant interpretations of the modeling differences, particularly with respect to intervention timing and uncertainty in parameter estimation.\\

\noindent
This paper is meticulously structured to offer a comprehensive understanding of our findings. Here is an overview of its organization: Section \ref{prelim} presents key probabilistic tools and foundational concepts. Section \ref{COVID-19} details the compartmental model and develops transition probability expressions for both stationary and non-stationary settings. Section \ref{example} presents simulations comparing the two regimes. Section \ref{conclusion} concludes with a discussion of practical implications and directions for future work.

\section{Preliminaries}\label{prelim}

\begin{definition}[Placing $n$ Indistinguishable Objects in $k$ Distinguishable Containers]
	
The number of possible ways of placing $n$ indistinguishable objects in $k$ distinguishable containers is given by:
	
		$$\begin{pmatrix}
			n+k-1 \\ k-1
		\end{pmatrix}
		=
		\begin{pmatrix}
			n+k-1 \\ n
		\end{pmatrix}
		=\dis\frac{(n+k-1)!}{n!(k-1)!}.
		$$

\noindent
Indeed, this is the number of combinations of $n+k-1$ things ($n$ of which are the unlabeled objects and $k-1$ are labeled containers), taken $n$ at a time (or similarly $k-1$ at a time).
\end{definition}

\begin{definition}[Binomial Probability Mass Function (PMF)]
	The Binomial PMF of a binomial random variable $X$ (that represents the number of successes in $n$ independent Bernoulli trials, each with a probability of success $p$), also denoted $\mathcal{B}(k;n,p)$ is given by:
	
		$$\P(X=k) 
		= \mathcal{B}(k;n,p)
		=\begin{pmatrix}
			n \\ k
		\end{pmatrix}
		p^k(1-p)^{n-k},
		$$
	
	\noindent
	where:
	\begin{itemize}
		\item $k$ is an integer representing the number of successes we are interested in,
		\item $\begin{pmatrix}	n \\ k	\end{pmatrix}$ is the  binomial coefficient representing the number of ways to choose $k$ successes out of $n$ trials,
		\item $p^k$ is the probability of getting $k$ successes,
		\item $(1-p)^{n-k}$ is the probability of getting $n-k$ failures.
	\end{itemize}
\end{definition}

\subsection{Chain-binomial Models}
The chain binomial models as a group assume that the generations of infectious are separated by a significant latent period and time of infectiousness. Thus, they are applicable to diseases in which cases or groups of cases are separated in time well enough to allow identification of successive generations of infection.

\begin{itemize}
\item Chain binomial models include the Reed-Frost model and the Greenwood model, which are two paradigms for modelling a disease	spread as discrete-time Markov chains.
\item They are called chain-binomial models because the transition probabilities are governed by binomial random variables.
\end{itemize}

\noindent
In an $SI$ model, we denote the generations of susceptible and infectious by $S$ and $I$ respectively. Let $N$ be the size of the population, $S_t$ the number of susceptible at time $t$, $I_t$ the number of infectious at time $t$, $\P(I_t)$ the probability that a susceptible becomes infected at time $t$. We have:
\begin{itemize}
	\item $S_0 + I_0 = N$: initial condition
	\item $S_{t+1}+I_{t+1} = S_t$ for times $t=0, 1, \cdots$: the infectious at time $t$ are removed from the process at time $t+1$.
	\item $S_t + \displaystyle\sum_{\tau=0}^{t}I_{\tau} = N, ~~ t=0, 1, \cdots$: at any time $t$, the the number of susceptible plus the number of infectious since time $0$ gives the total population.
	\item The number of infectious at time $t+1$ is given by the Binomial random variable of parameters $S_t$ and $\P(I_t)$. The corresponding transition probability is given by the $k^{th}$ component of the Binomial PMF as in the following definitions.
	
\end{itemize}

\begin{definition}[Greenwood Model]
	In the Greenwood model, the transition probability is given by:
	
		\begin{equation}\label{eq1}
			\mathbb{P}(I_{t+1}=k|S_t=x, I_t=y) 
			= \mathcal{B}(k;x,p) = 
			\left\{ 	
			\begin{array}{rllll}
				\begin{pmatrix}
					x \\ k
				\end{pmatrix}
				p^k(1-p)^{x-k} &\text{if}& 0\leq k \leq x \\
				0&\text{if}& k > x,
			\end{array}
			\right.
		\end{equation}
	
	\noindent
	where $k$ is an integer and $p$ is constant.
\end{definition}

\begin{rem}
	The Greenwood model is obtained if, instead of transfer by close contact, the transfer of infection occurs by contact of susceptibles with infectious material that is relatively widely spread, so that $p$, is a constant not depending on the number $y$ of infectious.
\end{rem}

\begin{definition}[Reed-Frost Model]
	In the Reed-Frost model, the transition probability is given by:
	
		\begin{equation}\label{eq2}
			\mathbb{P}(I_{t+1}=k|S_t=x, I_t=y) 
			= \mathcal{B}(k;x,p(y)) = 
			\left\{ 	
			\begin{array}{rllll}
				\begin{pmatrix}
					x \\ k
				\end{pmatrix}
				p(y)^k(1-p(y))^{x-k} &\text{if}& 0\leq k \leq x \\
				0&\text{if}& k > x,
			\end{array}
			\right.
		\end{equation}
	
	\noindent
	where $k$ is an integer and the probability any susceptible escapes being infected when there are $y$ infectious is $1-p(y) = (1- p(1))^y$, with $p(1)$ being the probability that a susceptible is infected by one given infectious, so that $p(y) = 1-(1- p(1))^y$.
\end{definition}

\begin{rem}
	The Reed-Frost model separates the probability of multiple contacts between a	susceptible and one infectious from the probability of multiple contacts of a susceptible with different infectious.
\end{rem}

\noindent
For additional information on Chain-binomial models, we refer to \cite{jacquez1987note, palopoli2023markovian} and the references therein.

\subsection{Markov Decision Process}
\begin{definition}[Markov Decision Process]\label{MDP}
	A Markov Decision Process (MDP) is a tuple $\langle \mathcal{S}, \mathcal{A}, \mathcal{P}, \mathcal{R}, \gamma\rangle$ where:
	\begin{itemize}
		\item $\gamma$ is a discount factor $\gamma\in [0,1]$
		\item $\mathcal{S}$ is a finite set of states
		\item $\mathcal{A}$ is a finite set of actions
		\item $\mathcal{P}$ is a state transition probability matrix, with the transition from a state $s$ to a successor state $s'$ defined by 
		$$\mathcal{P}_{ss'}^a=\P[S_{t+1}=s'|S_t=s, A_t=a].$$
		\item $\mathcal{R}$ is a reward function, with the reward for going from state $s$ to state $s'$ while taking action $a$ defined by $$R_{s}^a = \mathbb{E}[R_{t+1}|S_t=s, A_t=a].$$
	\end{itemize}
\end{definition}

\noindent
This study focuses on transition probabilities, which are crucial in MDP analysis for quantifying the likelihood of transitioning between states and enabling accurate modeling and optimization of decision-making processes. For more information on MDP, we refer to \cite{silver2015mdp}.

\section{A COVID-19 Scenario}\label{COVID-19}

\noindent
Moving forward, the model adopted will be a generalization of the Reed-Frost model. Indeed, Greenwood model seems not very convenient in our case, given that one of the protocol measure to control the spread of Covid-19 is social distancing, which helps avoid close contact with infectious individuals.

\subsection{Model Description}\label{SectDescript}
\begin{figure}[h!]
	\centering
	\includegraphics[height=6cm]{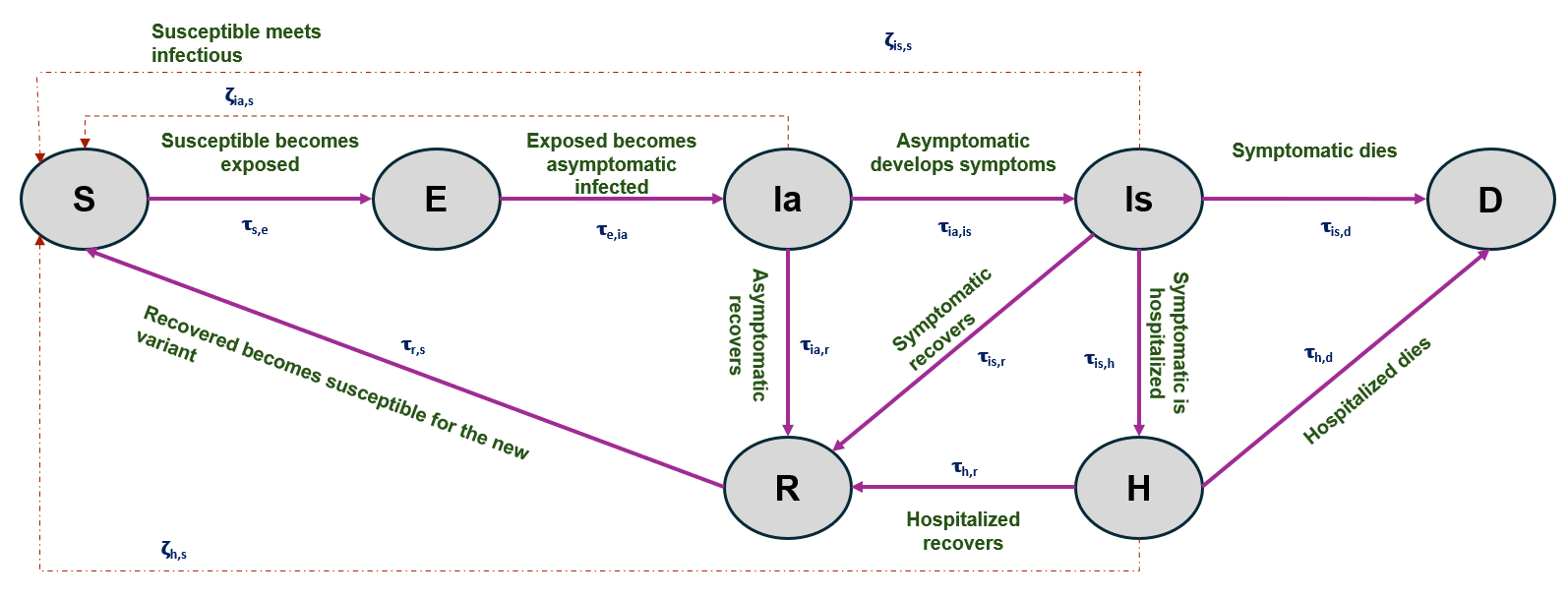}
	\caption{Dynamics of COVID-19 spread}
	\label{FigModel}
\end{figure}

We are interested in the following model (Figure \ref{FigModel}), that simulate the spread of COVID-19 through a population, divided into different compartments, each representing a distinct stage of the disease. 

\noindent
This model is typically represented using the following system of differential equations:

\begin{equation}\label{diff}
	\left\{
	\begin{array}{rlllll}
		\dis\frac{dS}{dt}&=& -\left(\zeta_{ia,s}I_a + \zeta_{is,s}I_s + \zeta_{h,s}H\right)\dis\frac{S}{N} + \tau_{r,s}R \\~\\
		\dis\frac{dE}{dt}&=& \left(\zeta_{ia,s}I_a + \zeta_{is,s}I_s + \zeta_{h,s}H\right)\dis\frac{S}{N} - \tau_{e,ia}E \\~\\
		\dis\frac{dI_a}{dt}&=& \tau_{e,ia}E - (\tau_{ia,is} + \tau_{ia,r})I_a \\~\\
		\dis\frac{dI_s}{dt}&=& \tau_{ia,is}I_a - (\tau_{is,r} + \tau_{is,h} + \tau_{is,d})I_s \\~\\
		\dis\frac{dH}{dt}&=& \tau_{is,h}I_s - (\tau_{h,r} + \tau_{h,d})H \\~\\
		\dis\frac{dR}{dt}&=& \tau_{is,r}I_s + \tau_{ia,r}I_a + \tau_{h,r}H - \tau_{r,s}R \\~\\
		\dis\frac{dD}{dt}&=& \tau_{h,d}H + \tau_{is,d}I_s
	\end{array}
	\right.,
\end{equation}
where $S, E, I_a, I_s, H, R$ and $D$ represent the states of Susceptible, Exposed, Asymptomatic Infected, Symptomatic Infected, Hospitalized, Recovered and Deceased individuals respectively.

\begin{enumerate}[(i)]
	\item Susceptible: Individuals who have not been infected with the virus but are at risk of becoming infected. They can move to the Exposed compartment upon contact with an infected individual.
	
	\item Exposed: Individuals who have been exposed to the virus and are in the incubation period. They are infected but not yet infectious. After the incubation period, they transition to the Asymptomatic Infected compartment.
	
	\item Asymptomatic Infected: Individuals who have been infected with the virus but do not show symptoms. They can still transmit the virus to Susceptible individuals. After a certain period, they either recover or transition to the symptomatic infected compartment.
	
	\item Symptomatic Infected: Individuals who show symptoms of the infection. These individuals can be further categorized by the severity of symptoms. Depending on the severity and progression of the disease, they may either recover, be hospitalized, or eventually transition to the Deceased compartment if the infection becomes fatal.
	
	\item Hospitalized: Individuals with severe symptoms who require intensive medical care. Their outcomes can vary: they may recover and move to the Recovered compartment or, in severe cases, may not survive and move to the Deceased compartment.
	
	\item Recovered: Individuals who have recovered from the infection but their immunity can last only for some period. They cannot transmit the disease but are still at risk of infection.
	
	\item Deceased: Individuals who have died due to the infection.
\end{enumerate}

\noindent
The transitions between compartments are governed by probabilities, which can be stationary (constant over time) or non-stationary (varying over time). These probabilities depend on various factors, including the disease's natural progression, intervention measures, and population behavior. The following probabilistic parameters are extremely important in studying the progression of the disease under consideration:
\begin{enumerate}[(a)]
	\item $\tau_{s,e}:$ probability for a susceptible individual to be exposed to the disease. It depends on the contact rate with infected individuals (asymptomatic, symptomatic or hospitalized), influenced by factors like social distancing, mask usage, vaccination rates, and so on.
	
	\item $\tau_{e,ia}:$ probability for an individual to be infected by the disease without exhibiting symptoms. This can be influenced by demographic factors such as underlying health conditions.
	
	\item $\tau_{ia,r}:$ probability for an individual to recover from asymptomatic infection. This probability is more likely based on the average duration of the infection.
	
	\item $\tau_{ia,is}:$ probability for an individual to develop symptoms subsequent to previous asymptomatic infection.
	
	\item Symptomatic individuals can have different pathways: mild cases may recover without hospitalization, severe cases may require hospitalization, followed by recovery or death, fatal cases transition to Deceased.
	
	\begin{itemize}
		\item $\tau_{is,h}:$ probability for an individual to require hospitalization in the Intensive Care Unit (ICU) due to complications arising from the symptoms of the disease.
		
		\item $\tau_{is,r}:$ probability for an individual to successfully recover after exhibiting symptoms of the disease.
		
		\item $\tau_{is,d}:$ probability for an individual to experience complications leading to death after displaying symptoms of the disease.
	\end{itemize}
	
	\item Hospitalized individuals may recover or succumb to the disease, with probabilities influenced by the quality of healthcare, the severity of the disease, and comorbidities.
	
	\begin{itemize}
		\item $\tau_{h,d}:$ probability for an individual to succumb to mortality following hospitalization in the ICU.
		
		\item $\tau_{h,r}:$ probability for an individual to successfully recover after being discharged from the ICU.
	\end{itemize}
	
	\item $\tau_{r,s}:$ probability for an individual to become susceptible again after recovering from the disease.
	
	\item $\zeta_{ia,s}:$ probability to contract the infection in one meeting with an individual from $I_a$.
	
	\item $\zeta_{is,s}:$ probability to contract the infection in one meeting with an individual from $I_s$.
	
	\item $\zeta_{h,s}:$ probability to contract the infection in one meeting with an individual from $H$.
\end{enumerate}

\begin{rem}
	We note that susceptible individuals become exposed by coming into contact with asymptomatically infectious, symptomatically infectious and hospitalized individuals.
\end{rem}

\noindent
The only deterministic parameter, denoted $N$, represents the total population in the considered region.

\subsection{Balanced System}\label{SectBalance}
Let's consider the following diagram, representing the transitions between individuals from a state to another.

\begin{figure}[h!]
	\centering
	\includegraphics[height=5cm]{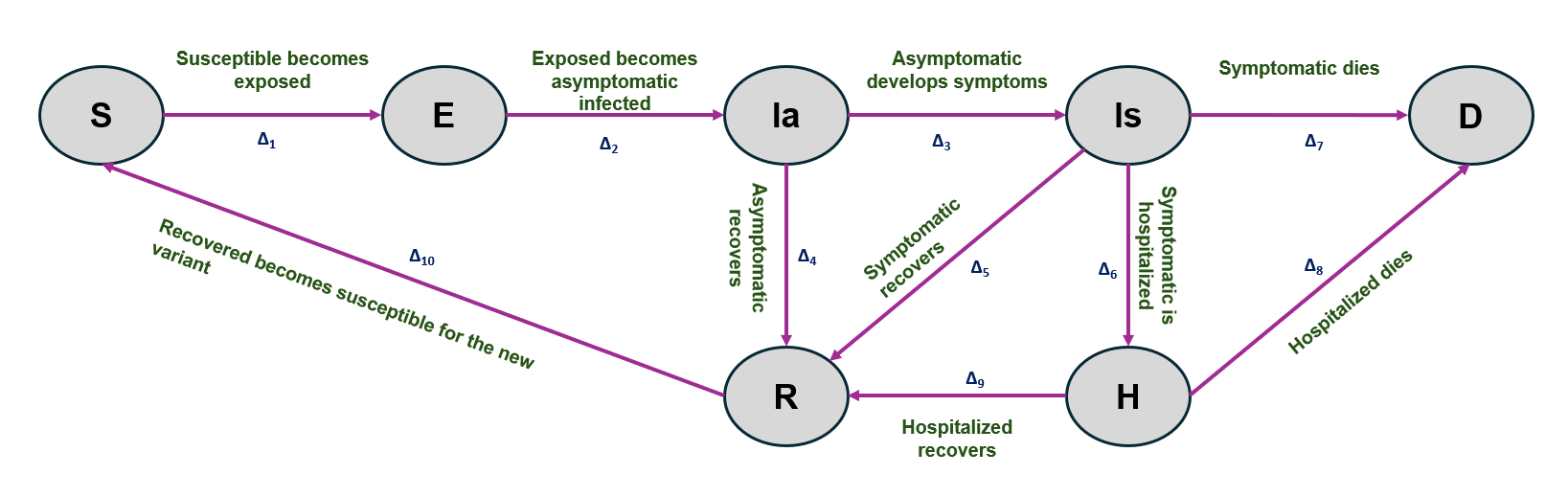}
	\caption{Balanced system}
	\label{FigBalance}
\end{figure}

\noindent
The number of people in each state determines the evolution of the population and our system can be conveniently modelled as a discrete-time Markov chain. The state of the Markov chain will be associated with the septuple collected in the following vector
\begin{equation}
	V_k= [|S|_k, |E|_k, |I_a|_k, |I_s|_k, |H|_k, |R|_k, |D|_k]^T,
\end{equation}
where $|S|_k, |E|_k, |I_a|_k, |I_s|_k, |H|_k, |R|_k$, and $|D|_k$ represent the cardinalities of $S, E, I_a, I_s, H, R$, and $D$ respectively at stage $k$, with
\begin{equation}
	|S|_k + |E|_k + |I_a|_k + |I_s|_k + |H|_k + |R|_k + |D|_k = N.
\end{equation}

\noindent
In the main section of this work, we will show how to compute the transition probability:
$$\P (V_{k+1}=v'|V_k=v),$$
where $v$ and $v'$ are respectively the state vector values before and after the transition, characterized by the following septuple represents

$$v'-v = [\Delta_{|S|}, \Delta_{|E|}, \Delta_{|I_a|}, \Delta_{|I_s|}, \Delta_{|H|}, \Delta_{|R|}, \Delta_{|D|}]^T.$$

\noindent
Consider also the vector
\begin{equation}
	\Delta = [\Delta_1, \Delta_2, \Delta_3, \Delta_4, \Delta_5, \Delta_6, \Delta_7, \Delta_8, \Delta_9, \Delta_{10}]^T,
\end{equation}
where for each $i=1,2,\cdots,10$, $\Delta_i$ represents the flow of individuals between different states, with $\Delta_i \in [0,N]\cap \N$ satisfying the following balanced equation:

\begin{equation}\label{SystBal}
	\left\{
	\begin{array}{rlllll}
		B_1:& \Delta_{|S|} &=& \Delta_{10} - \Delta_1  \\
		B_2:& \Delta_{|E|} &=& \Delta_1 - \Delta_2  \\
		B_3:& \Delta_{|I_a|} &=& \Delta_2 - \Delta_3 - \Delta_4  \\
		B_4:& \Delta_{|I_s|} &=& \Delta_3 - \Delta_5 - \Delta_6 - \Delta_7  \\
		B_5:& \Delta_{|D|} &=& \Delta_7 + \Delta_8  \\
		B_6:& \Delta_{|H|} &=& \Delta_6 - \Delta_8 - \Delta_9  \\
		B_7:& \Delta_{|R|} &=& \Delta_4 + \Delta_5 + \Delta_9 - \Delta_{10}.  \\
	\end{array}
	\right.
\end{equation}

\noindent
The flows from different states are subjected to the following constraints:
\begin{equation}\label{Constr}
	\left\{
	\begin{array}{rlllll}
		|S|_k & \geq & \Delta_1 \\
		|E|_k & \geq & \Delta_2 \\
		|I_a|_k & \geq & \Delta_3 + \Delta_4 \\
		|I_s|_k & \geq & \Delta_5 + \Delta_6 + \Delta_7 \\
		|H|_k & \geq & \Delta_8 + \Delta_9 \\
		|R|_k & \geq & \Delta_{10} \\
	\end{array}
	\right.
\end{equation}
This ensures that the number of subjects remaining in the compartments $S, E, I_a, I_s, H$ and $R$ at stage $k$ remains non-negative.\\

\noindent
Now, denoting by $l(\cdot)$ an assignment of variables: $\Delta_i = \delta_i$ for $i=1,\cdots,10$, let's introduce the following notations:

$$l_1: (\Delta_1 = \delta_1)$$
is a function linking the variable $\Delta_1$, defined via the balance equation $B_1$.

$$l_2: (\Delta_2 = \delta_1 - \Delta_{|E|})$$
is a function linking the variable $\Delta_2$, defined via $l_1$ and the balance equation $B_2$.

$$l_3: (\Delta_3=\delta_3, \Delta_4 = \delta_1-\Delta_{|E|}-\delta_3-\Delta_{|I_a|})$$
is a function linking the variables $\Delta_3$ and $\Delta_4$, defined via $l_2$ and the balance equation $B_3$.

$$l_4: (\Delta_3 = \delta_3, \Delta_5 = \delta_5, \Delta_6 = \delta_6, \Delta_7 = \delta_3 - \delta_5 - \delta_6 - \Delta_{|I_s|})$$
is a function linking the variables $\Delta_3, \Delta_5, \Delta_6$ and $\Delta_7$, defined via $l_3$ and the balance equation $B_4$.

$$l_5: (\Delta_8 = \Delta_{|D|} - \delta_3 + \delta_5 + \delta_6 + \Delta_{|I_s|}, \Delta_9 = \delta_3 - \Delta_{|H|} - \Delta_{|D|} -\delta_5 - \Delta_{|I_s|})$$
is a function linking the variables $\Delta_8$ and $\Delta_9$, defined via $l_4$ and the balance equations $B_5$ and $B_6$.

$$l_6: (\Delta_{10} = \delta_1 - \Delta_{|E|} - \Delta_{|I_a|} - \Delta_{|H|} - \Delta_{|D|} - \Delta_{|I_s|} - \Delta_{|R|})$$
is a function linking the variable $\Delta_{10}$, defined via $l_3, l_4, l_5$ and the balance equation $B_7$.\\

\noindent
In this study, the stationary and non-stationary transition probabilities will be critical for understanding how different interventions (like vaccination, social distancing, hand sanitization, masking, testing and contact tracing, ICU, quarantine) affect the progression of the disease. By modeling these probabilities accurately, policymakers can predict outcomes under various scenarios and make informed decisions about resource allocation, healthcare planning, and economic impacts.

\subsection{Stationary Transition Probabilities}\label{SectStation}
Stationary transition probabilities assume that the probabilities of transitioning between states remain constant over time. For diseases where the transition rates are relatively stable over time, stationary models can provide reliable long-term predictions and insights. The models are often mathematically simpler and easier to analyze and allow for the use of established techniques and tools in epidemiological modeling. In such cases, disease dynamics, treatment effects, and other relevant factors are assumed to be stable and unchanging over time. \\

\noindent Let $\delta_i, \delta_j, \cdots$ represent the flow of individuals moving from one state. Consider $\V_{\delta_i, \delta_j, \cdots}$ as the event where the flow of individuals is such that $\delta_i$ transitions to a new state, $\delta_j$ transitions to another state, and so forth, as specified. That is, $\V_{\delta_i, \delta_j, \cdots}=$ "exactly $\delta_i$ individuals move to one state and $\delta_j$ individuals to another state and so on, given $V_k$", where $V_k$ is defined by \eqref{Vk}.

\begin{theo}[Probability of Leaving States]\label{Theoprobleav}~\\
	Considering the septuple 
	\begin{equation}\label{Vk}
	      V_k = [ |S|_k, |E|_k, |I_a|_k, |I_s|_k, |H|_k, |R|_k, |D|_k ]^T
	\end{equation}
	where each entry represent the cardinality of the corresponding compartment at epoch $k$, the probabilities of leaving state $S, E, I_a, I_s, H$ and $R$ are respectively given by the following items.
	
	\begin{enumerate}[(i)]
		\item 
		\begin{equation}\label{ProbLeavS}
		\P(\V_{\delta_1})=
		\left\{
		\begin{array}{llll}
			0 &\text{if}& \delta_1 > |S|_k \\
			\begin{pmatrix}
				|S|_k \\~\\ \delta_1
			\end{pmatrix}
			(\P(g_k))^{\delta_1}(1-\P(g_k))^{|S|_k-\delta_1} &\text{if}& \delta_1 \leq |S|_k
		\end{array}
		\right.,
		\end{equation}
		where 
		$$\P(g_k)_M = 1- \left( 1 - \dis\frac{(\zeta_{ia,s}+\zeta_{is,s}+\zeta_{h,s})(|E|_k + |I_a|_k+|I_s|_k+|H|_k)}{N-|D|_k} \right)^M.
		$$
		with $M$ representing the number of meeting and $g_k (\overline{g}_k)$ the event: "an individual (does not) become infected after meeting an infectious person, given $V_k$".
		
		\item 
		\begin{equation}\label{ProbLeavE}
			\P(\V_{\delta_2})=
			\left\{
			\begin{array}{llll}
				0 &\text{if}& \delta_2 > |E|_k \\
				\begin{pmatrix}
					|E|_k \\~\\ \delta_2
				\end{pmatrix}
				\tau_{e,ia}^{\delta_2}(1-\tau_{e,ia})^{|E|_k-\delta_2} &\text{if}& \delta_2 \leq |E|_k
			\end{array}
			\right.
		\end{equation}
		
		\item 
		\begin{equation}\label{ProbLeavIa}
			\P(\V_{\delta_3,\delta_4})=
			\left\{
			\begin{array}{llll}
				0 &\text{if}& \delta_3+\delta_4 > |I_a|_k \\
				\M_{|I_a|_k,\delta_3,\delta_4}\tau_{ia,is}^{\delta_3}\tau_{ia,r}^{\delta_4}(1-(\tau_{ia,is}+\tau_{ia,r}))^{|I_a|_k-(\delta_3+\delta_4)} &\text{if}& \delta_3+\delta_4 \leq |I_a|_k
			\end{array}
			\right.
		\end{equation}
		
		\item 
		\begin{equation}\label{ProbLeavIs}
			\P(\V_{\delta_5,\delta_6, \delta_7})=
			\left\{
			\begin{array}{llll}
				0 ~~~\text{if}~~~ \delta_5+\delta_6+\delta_7 > |I_s|_k  \\~\\
				\M_{|I_s|_k,\delta_5,\delta_6, \delta_7}\tau_{is,r}^{\delta_5}\tau_{is,h}^{\delta_6}\tau_{is,d}^{\delta_7}(1-(\tau_{is,r}+\tau_{is,h}+\tau_{is,d}))^{|I_s|_k-(\delta_5+\delta_6+\delta_7)}\\
				 \text{if}~~ \delta_5+\delta_6+\delta_7 \leq |I_s|_k
			\end{array}
			\right.
		\end{equation}
		
		\item 
		\begin{equation}\label{ProbLeavH}
			\P(\V_{\delta_8,\delta_9})=
			\left\{
			\begin{array}{llll}
				0 &\text{if}& \delta_8+\delta_9 > |H|_k \\
				\M_{|H|_k,\delta_8,\delta_9}\tau_{h,d}^{\delta_8}\tau_{h,r}^{\delta_9}(1-(\tau_{h,d}+\tau_{h,r}))^{|H|_k-(\delta_8+\delta_9)} &\text{if}& \delta_8+\delta_9 \leq |H|_k
			\end{array}
			\right.
		\end{equation}

        \item 
        \begin{equation}\label{ProbLeavR}
            \P(\V_{\delta_{10}})=1,
        \end{equation}
        considering that no individual gains immunity after recovering from the disease.
		
	\end{enumerate}
\end{theo}

\noindent
\begin{proof}
\begin{enumerate}[(i)]
		\item \textbf{Probability of Leaving State $S$}\\
	Let's consider the events:\\
	$c_k (\overline{c}_k):$ "an individual meets one person and (does not) contract the infection, given $V_k$".
	
	\noindent
	$m_k (\overline{m}_k)$: "an individual meets one person who is (not) infectious, given $V_k$".
	
	\noindent
	$g_k (\overline{g}_k)$: "an individual (does not) become infected after meeting an infectious person, given $V_k$".
	
	\noindent
	Suppose $M=$ number of meetings allowed in each period.
	
	\noindent
	We recall also that $\zeta_{ia,s}, \zeta_{is,s}, \zeta_{h,s} =$ probability to contract the infection in one meeting with an individual who is asymptomatic, symptomatic or hospitalized respectively.
	
	\noindent
	We can write:
	$$g_k = \left\{ p\in S_k \wedge p\in E_{k+1} \right\},$$
	where $p$ is a generic person.
	
	\noindent
	We have:
	$$\P(g_k)_M = 1- (\P(\overline{c}_k))^M$$
	and
	$$\overline{c}_k = \overline{m}_k \vee (m_k \wedge \overline{g}_k).$$
	Then:
	
	$$
	\begin{array}{rlllll}
		\P(\overline{c}_k) &=& \P (\overline{m}_k \vee (m_k \wedge \overline{g}_k))\\~\\
		&=&\P(\overline{m}_k) + \P (m_k \wedge \overline{g}_k)\\~\\
		&=& \P(\overline{m}_k) + \P(\overline{g}_k|m_k)\P(m_k).
	\end{array}
	$$
	We know that
	$$\P(\overline{g}_k|m_k) = 1 - (\zeta_{ia,s}+\zeta_{is,s}+\zeta_{h,s}),$$
	Hence:
	
	\noindent
	$$
	\begin{array}{rlllll}
		\P(\overline{c}_k) &=& \P(\overline{m}_k) + (1 - (\zeta_{ia,s}+\zeta_{is,s}+\zeta_{h,s}))\P(m_k)\\~\\
		&=& \dis\frac{|S|_k + |R|_k}{N-|D|_k} + (1 -(\zeta_{ia,s}+\zeta_{is,s}+\zeta_{h,s}))\dis\frac{|E|_k + |I_a|_k+|I_s|_k+|H|_k}{N-|D|_k} \\~\\
		&=& \dis\frac{|S|_k + |R|_k + |E|_k + |I_a|_k+|I_s|_k+|H|_k}{N-|D|_k} \\~\\
		&&- (\zeta_{ia,s}+\zeta_{is,s}+\zeta_{h,s})\dis\frac{|E|_k + |I_a|_k+|I_s|_k+|H|_k}{N-|D|_k} \\~\\
		&=& 1 - \dis\frac{(\zeta_{ia,s}+\zeta_{is,s}+\zeta_{h,s})(|E|_k + |I_a|_k+|I_s|_k+|H|_k)}{N-|D|_k}.
	\end{array}
	$$
	
	\noindent
	Therefore,
	
	$$\P(g_k)_M = 1- \left( 1 - \dis\frac{(\zeta_{ia,s}+\zeta_{is,s}+\zeta_{h,s})(|E|_k + |I_a|_k+|I_s|_k+|H|_k)}{N-|D|_k} \right)^M.
	$$
	
	\noindent
	Now, let's consider the event $\V_{\delta_1}:$ "exactly $\delta_1$ susceptible individuals are exposed to the disease, given $V_k$".
	
	\begin{itemize}
		\item If $\delta_1 > |S|_k$, then $\P(\V_{\delta_1})=0$.
		\item If $\delta_1 \leq |S|_k$ we have:
		$$\P(\V_{\delta_1}) = \mathcal{B}(\delta_1; |S|_k, \P(g_k))=
		\begin{pmatrix}
			|S|_k \\~\\ \delta_1
		\end{pmatrix}
		(\P(g_k))^{\delta_1}(1-\P(g_k))^{|S|_k-\delta_1},$$
		where the binomial 
		$
		\begin{pmatrix}
			|S|_k \\~\\ \delta_1
		\end{pmatrix}
		$ represents the possible combinations of $\delta_1$ individuals out of $|S|_k$ and $\P(g_k)$ is the probability for an individual to become infected after meeting an infectious person (in one meeting).
	\end{itemize}
	
	\item \textbf{Probability of Leaving State $E$}\\
	Let's consider the event
	$\V_{\delta_2}:$ "exactly $\delta_2$ individuals are infected by the disease without exhibiting symptoms, given $V_k$".
	
	\begin{itemize}
		\item If $\delta_2 > |E|_k$, then $\P(\V_{\delta_2})=0$.
		\item If $\delta_2 \leq |E|_k$ we have:
		$$\P(\V_{\delta_2}) = \mathcal{B}(\delta_2; |E|_k, \tau_{e,ia})=
		\begin{pmatrix}
			|E|_k \\~\\ \delta_2
		\end{pmatrix}
		\tau_{e,ia}^{\delta_2}(1-\tau_{e,ia})^{|E|_k-\delta_2}.
		$$
	\end{itemize}
	
	\item \textbf{Probability of Leaving State $I_a$}\\
	Let's consider the event $\V_{\delta_3,\delta_4}$: "exactly $\delta_3$ individuals develop symptoms subsequent to previous asymptomatic infection and $\delta_4$ individuals recover from asymptomatic infection, given $V_k$".
	
	\begin{itemize}
		\item If $\delta_3+\delta_4 > |I_a|_k$, then $\P(\V_{\delta_3,\delta_4})=0$.
		\item If $\delta_3+\delta_4 \leq |I_a|_k$ we have:
		$$\P(\V_{\delta_3,\delta_4}) = \M_{|I_a|_k,\delta_3,\delta_4}\tau_{ia,is}^{\delta_3}\tau_{ia,r}^{\delta_4}(1-(\tau_{ia,is}+\tau_{ia,r}))^{|I_a|_k-(\delta_3+\delta_4)},
		$$
		where the multinomial coefficient $$\M_{|I_a|_k,\delta_3,\delta_4}=
		\begin{pmatrix}
			|I_a|_k \\~\\ 
			\delta_3,\delta_4, |I_a|_k-(\delta_3+\delta_4)
		\end{pmatrix}
		$$
		represents the possible combinations of $\delta_3+\delta_4$ individuals out of $|I_a|_k$.
	\end{itemize}
	
	\item \textbf{Probability of Leaving State $I_s$}\\
	Let's consider the event $\V_{\delta_5,\delta_6,\delta_7}$: "exactly $\delta_5$ individuals successfully recover after exhibiting symptoms and $\delta_6$ individuals require ICU hospitalization due to complications arising from the symptoms and $\delta_7$ individuals experience complications leading to death after displaying symptoms of the disease, given $V_k$".
	
	\begin{itemize}
		\item If $\delta_5+\delta_6+\delta_7 > |I_s|_k$, then $\P(\V_{\delta_5,\delta_6, \delta_7})=0$.
		\item If $\delta_5+\delta_6+\delta_7 \leq |I_s|_k$ we have:
		$$\P(\V_{\delta_5,\delta_6, \delta_7}) = \M_{|I_s|_k,\delta_5,\delta_6, \delta_7}\tau_{is,r}^{\delta_5}\tau_{is,h}^{\delta_6}\tau_{is,d}^{\delta_7}(1-(\tau_{is,r}+\tau_{is,h}+\tau_{is,d}))^{|I_s|_k-(\delta_5+\delta_6+\delta_7)}.
		$$
	\end{itemize}
	
	\item \textbf{Probability of Leaving State $H$}\\
		Let's consider the event $\V_{\delta_8,\delta_9}$: "exactly $\delta_8$ individuals succumb to mortality following hospitalization in the ICU and $\delta_9$ individuals successfully recover after being discharged from the ICU, given $V_k$".
	
		\begin{itemize}
		\item If $\delta_8+\delta_9 > |H|_k$, then $\P(\V_{\delta_8,\delta_9})=0$.
		\item If $\delta_8+\delta_9 \leq |H|_k$ we have:
		$$\P(\V_{\delta_8,\delta_9}) = \M_{|H|_k,\delta_8,\delta_9}\tau_{h,d}^{\delta_8}\tau_{h,r}^{\delta_9}(1-(\tau_{h,d}+\tau_{h,r}))^{|H|_k-(\delta_8+\delta_9)}.
		$$
		\end{itemize}
		
	\item \textbf{Probability of Leaving State $R$}\\
		Let's consider the event $\V_{\delta_{10}}$: "exactly $\delta_{10}$ individuals become susceptible again after recovering from the disease, given $V_k$".
        Given that no individual gains immunity after recovering from the disease, there is a hundred percent likelihood of becoming susceptible again. Therefore, $\P(\V_{\delta}) = \tau_{r,s} = 1$.
\end{enumerate}
This completes the proof of Theorem \ref{Theoprobleav}.
\end{proof}

\noindent
At this point, we have all we need to proceed with the different transition probabilities.

\begin{theo}[Transition Probabilities]\label{Theoprobtransit}~\\
	The transition probabilities from a state to another are given by the following, where
        $$\beta = 1 - \tau_{is,r} - \tau_{is,h} - \tau_{is,d}, ~~ Q_1 = |I_a|_k - \delta_3 - \delta_4, ~\text{and} ~ Q_2 = |I_s|_k - \delta_5 - \delta_6 - \delta_7.$$
        
	\begin{enumerate}[(a)]
		\item \textbf{Transition Probability From State $S$ to State $E$}
		\begin{equation}\label{StoE}
			\begin{array}{rllll}
				\P(E|S) &=& \begin{pmatrix}
					|S|_k \\~\\ \delta_1
				\end{pmatrix}
				\left(\dis\frac{(\zeta_{ia,s}+\zeta_{is,s}+\zeta_{h,s})(|E|_k + |I_a|_k+|I_s|_k+|H|_k)}{N-|D|_k}\right)^{\delta_1}\\~\\
				&&\left(1-\dis\frac{(\zeta_{ia,s}+\zeta_{is,s}+\zeta_{h,s})(|E|_k + |I_a|_k+|I_s|_k+|H|_k)}{N-|D|_k}\right)^{|S|_k-\delta_1}
			\end{array}
		\end{equation}
		
		\item \textbf{Transition Probability From State $E$ to State $I_a$}
		\begin{equation}\label{EtoIa}
			\P(I_a|E) = \begin{pmatrix}
				|E|_k \\~\\ \delta_2
			\end{pmatrix}
			\tau_{e,ia}^{\delta_2}(1-\tau_{e,ia})^{|E|_k-\delta_2}.
		\end{equation}
		
		\item \textbf{Transition Probability From State $I_a$ to State $I_s$}
		\begin{equation}\label{IatoIs}
			\P(I_s \mid I_a) = 
		\frac{
			\displaystyle\sum_{\delta_4 = 0}^{|I_a|_k - \delta_3} 
			\binom{|I_a|_k}{\delta_3, \delta_4, Q_1} 
			\tau_{ia,is}^{\delta_3} 
			\tau_{ia,r}^{\delta_4} 
			(1 - \tau_{ia,is} - \tau_{ia,r})^{Q_1}
		}{
			\dis\sum_{\delta_3 = 0}^{|I_a|_k}
			\sum_{\delta_4 = 0}^{|I_a|_k - \delta_3}
			\binom{|I_a|_k}{\delta_3, \delta_4, Q_1} 
			\tau_{ia,is}^{\delta_3} 
			\tau_{ia,r}^{\delta_4} 
			(1 - \tau_{ia,is} - \tau_{ia,r})^{Q_1}
		}.
		\end{equation}

		\item \textbf{Transition Probability From State $I_a$ to State $R$}
		\begin{equation}\label{IatoR}
			\P(R \mid I_a) = 
	\frac{
		\displaystyle\sum_{\delta_3 = 0}^{|I_a|_k - \delta_4} 
		\binom{|I_a|_k}{\delta_3, \delta_4, Q_1} 
		\tau_{ia,is}^{\delta_3} 
		\tau_{ia,r}^{\delta_4} 
		(1 - \tau_{ia,is} - \tau_{ia,r})^{Q_1}
	}{
		\dis\sum_{\delta_4 = 0}^{|I_a|_k}
		\sum_{\delta_3 = 0}^{|I_a|_k - \delta_4}
		\binom{|I_a|_k}{\delta_3, \delta_4, Q_1} 
		\tau_{ia,is}^{\delta_3} 
		\tau_{ia,r}^{\delta_4} 
		(1 - \tau_{ia,is} - \tau_{ia,r})^{Q_1}
	}.
		\end{equation}

		\item \textbf{Transition Probability From State $I_s$ to State $R$}
		\begin{equation}\label{IstoR}
			\P (R|I_s) = \frac{
		\displaystyle\sum_{\delta_6 = 0}^{|I_s|_k - \delta_5}\sum_{\delta_7 = 0}^{|I_s|_k - \delta_5-\delta_6} 
		\binom{|I_s|_k}{\delta_5, \delta_6, \delta_7, Q_2} 
		\tau_{is,r}^{\delta_5} 
		\tau_{is,h}^{\delta_6}\tau_{is,d}^{\delta_7} 
		\beta^{Q_2}
	}{
		\dis\sum_{\delta_5 = 0}^{|I_s|_k}
		\sum_{\delta_6 = 0}^{|I_s|_k - \delta_5}\sum_{\delta_7 = 0}^{|I_s|_k - \delta_5 - \delta_6}
		\binom{|I_s|_k}{\delta_5, \delta_6, \delta_7, Q_2} 
		\tau_{is,r}^{\delta_5} 
		\tau_{is,h}^{\delta_6}\tau_{is,d}^{\delta_7} 
		\beta^{Q_2}
	}.
		\end{equation}

        \noindent

		\item \textbf{Transition Probability From State $I_s$ to State $H$}
		\begin{equation}\label{IstoH}
			\P (H|I_s) = \frac{
		\displaystyle\sum_{\delta_5 = 0}^{|I_s|_k - \delta_6}\sum_{\delta_7 = 0}^{|I_s|_k - \delta_5-\delta_6} 
		\binom{|I_s|_k}{\delta_5, \delta_6, \delta_7, Q_2} 
		\tau_{is,r}^{\delta_5} \tau_{is,h}^{\delta_6}\tau_{is,d}^{\delta_7} 
		\beta^{Q_2}
	}{
		\dis\sum_{\delta_6 = 0}^{|I_s|_k}
		\sum_{\delta_5 = 0}^{|I_s|_k - \delta_6}\sum_{\delta_7 = 0}^{|I_s|_k - \delta_5 - \delta_6}
		\binom{|I_s|_k}{\delta_5, \delta_6, \delta_7, Q_2} 
		\tau_{is,r}^{\delta_5} \tau_{is,h}^{\delta_6}\tau_{is,d}^{\delta_7} 
		\beta^{Q_2}
	}.
		\end{equation}

		\item \textbf{Transition Probability From State $I_s$ to State $D$}
		\begin{equation}\label{IstoD}
			\P (D|I_s) = \frac{
		\displaystyle\sum_{\delta_5 = 0}^{|I_s|_k - \delta_7}\sum_{\delta_6 = 0}^{|I_s|_k - \delta_5-\delta_7} 
		\binom{|I_s|_k}{\delta_5, \delta_6, \delta_7, Q_2} 
		\tau_{is,r}^{\delta_5} \tau_{is,h}^{\delta_6}\tau_{is,d}^{\delta_7} 
		\beta^{Q_2}
	}{
		\dis\sum_{\delta_7 = 0}^{|I_s|_k}
		\sum_{\delta_5 = 0}^{|I_s|_k - \delta_7}\sum_{\delta_6 = 0}^{|I_s|_k - \delta_5 - \delta_7}
		\binom{|I_s|_k}{\delta_5, \delta_6, \delta_7, Q_2} 
		\tau_{is,r}^{\delta_5} \tau_{is,h}^{\delta_6}\tau_{is,d}^{\delta_7} 
		\beta^{Q_2}
	}.
		\end{equation}

		\item \textbf{Transition Probability From State $H$ to State $D$}
		\begin{equation}\label{HtoD}
			\P (D|H) = \frac{
		\displaystyle\sum_{\delta_9 = 0}^{|H|_k - \delta_8} 
		\binom{|H|_k}{\delta_8, \delta_9, |H|_k - \delta_8 - \delta_9} 
		\tau_{h,r}^{\delta_9} 
		\tau_{h,d}^{\delta_8} 
		(1 - \tau_{h,r} - \tau_{h,d})^{|H|_k - \delta_8 - \delta_9}
	}{
		\dis\sum_{\delta_8 = 0}^{|H|_k}
		\sum_{\delta_9 = 0}^{|H|_k - \delta_8}
		\binom{|H|_k}{\delta_8, \delta_9, |H|_k - \delta_8 - \delta_9} 
		\tau_{h,r}^{\delta_9} 
		\tau_{h,d}^{\delta_8} 
		(1 - \tau_{h,r} - \tau_{h,d})^{|H|_k - \delta_8 - \delta_9}
	}.
		\end{equation}

		\item \textbf{Transition Probability From State $H$ to State $R$}
		\begin{equation}\label{HtoR}
			\P (R|H) = \frac{
		\displaystyle\sum_{\delta_8 = 0}^{|H|_k - \delta_9} 
		\binom{|H|_k}{\delta_8, \delta_9, |H|_k - \delta_8 - \delta_9} 
		\tau_{h,r}^{\delta_9} 
		\tau_{h,d}^{\delta_8} 
		(1 - \tau_{h,r} - \tau_{h,d})^{|H|_k - \delta_8 - \delta_9}
	}{
		\dis\sum_{\delta_9 = 0}^{|H|_k}
		\sum_{\delta_8 = 0}^{|H|_k - \delta_9}
		\binom{|H|_k}{\delta_8, \delta_9, |H|_k - \delta_8 - \delta_9} 
		\tau_{h,r}^{\delta_9} 
		\tau_{h,d}^{\delta_8} 
		(1 - \tau_{h,r} - \tau_{h,d})^{|H|_k - \delta_8 - \delta_9}
	}.
		\end{equation}

		\item \textbf{Transition Probability From State $R$ to State $S$}
        \begin{equation}\label{RtoS}
			\P(S|R) = 1,
		\end{equation}
  considering that no individual gains immunity after recovering from the disease.
		
	\end{enumerate}
\end{theo}

\noindent
\begin{proof}
\begin{enumerate}[(a)]
	\item The probability of landing in State $E$ after leaving State $S$ is as follows:
		$$\begin{array}{rllll}
			\P(E|S) &=& \P(\V_{\delta_1}) \\~\\
			&=& \mathcal{B}(\delta_1; |S|_k, \P(g_k)) \\~\\
			&=& \begin{pmatrix}
				|S|_k \\~\\ \delta_1
			\end{pmatrix}(\P(g_k))^{\delta_1}(1-\P(g_k))^{|S|_k-\delta_1}.
		\end{array}
		$$
		
		\noindent
		Therefore, the transition probability from state $S$ to state $E$ is given by \eqref{StoE}.
		
	\item The probability of landing in State $I_a$ after leaving State $E$ is as follows:
		$$\begin{array}{rllll}
			\P(I_a|E) &=& \P(\V_{\delta_2}) \\~\\
			&=& \mathcal{B}(\delta_2; |E|_k, \tau_{e,ia}).
		\end{array}
		$$
		
		\noindent
		Therefore, the transition probability from state $E$ to state $I_a$  is given by \eqref{EtoIa}.

	\item The probability of landing in State $I_s$ after leaving State $I_a$ is as follows:
	$$\P (I_s|I_a) = \dis\frac{\P(I_s\cap I_a)}{\P(I_a)},$$
	where 
\begin{equation}
	\P (I_s\cap I_a) = \displaystyle\sum_{\delta_4 = 0}^{|I_a|_k - \delta_3} 
	\binom{|I_a|_k}{\delta_3, \delta_4, Q_1} 
	\tau_{ia,is}^{\delta_3} 
	\tau_{ia,r}^{\delta_4} 
	(1 - \tau_{ia,is} - \tau_{ia,r})^{Q_1}
\end{equation}
is obtained by summing over all possible values of $\delta_4$, and

\begin{equation}
	\P(I_a) = \sum_{\delta_3 = 0}^{|I_a|_k}
	\sum_{\delta_4 = 0}^{|I_a|_k - \delta_3}
	\binom{|I_a|_k}{\delta_3, \delta_4, Q_1} 
	\tau_{ia,is}^{\delta_3} 
	\tau_{ia,r}^{\delta_4} 
	(1 - \tau_{ia,is} - \tau_{ia,r})^{Q_1}.
\end{equation}

\noindent
Therefore, the transition probability from state $I_a$ to state $I_s$  is given by \eqref{IatoIs}.

	\item The probability of landing in State $R$ after leaving State $I_a$ is as follows:
	$$\P (R|I_a) = \dis\frac{\P(R\cap I_a)}{\P(I_a)},$$
	where
\begin{equation}
	\P (R\cap I_a) = \displaystyle\sum_{\delta_3 = 0}^{|I_a|_k - \delta_4} 
	\binom{|I_a|_k}{\delta_3, \delta_4, Q_1} 
	\tau_{ia,is}^{\delta_3} 
	\tau_{ia,r}^{\delta_4} 
	(1 - \tau_{ia,is} - \tau_{ia,r})^{Q_1}
\end{equation}
is obtained by summing over all possible values of $\delta_3$, and

\begin{equation}
	\P(I_a) = \sum_{\delta_4 = 0}^{|I_a|_k}
	\sum_{\delta_3 = 0}^{|I_a|_k - \delta_4}
	\binom{|I_a|_k}{\delta_3, \delta_4, Q_1} 
	\tau_{ia,is}^{\delta_3} 
	\tau_{ia,r}^{\delta_4} 
	(1 - \tau_{ia,is} - \tau_{ia,r})^{Q_1}.
\end{equation}

\noindent
Therefore, the transition probability from state $I_a$ to state $R$  is given by \eqref{IatoR}.
\end{enumerate}

\begin{rem}
    The transition probabilities \eqref{IstoR}-\eqref{HtoR} are obtained following the same reasoning as in Part (c) and (d).
\end{rem}

\begin{rem}
    Given that no individual gains immunity after recovering from the disease, there is hundred percent chance of landing in $S$ after leaving $R$. Hence,
        $$\begin{array}{rllll}
			\P(S|R) &=& 1.
		\end{array}$$
\end{rem}

\noindent
This completes the proof of Theorem \ref{Theoprobtransit}.
\end{proof}\\

\noindent Stationary transition probabilities assume that the process reaches a stable equilibrium where the disease dynamics are predictable. This model is not a good fit for predicting the long term dynamics of Covid-19 due to ongoing uncertainties such as: the evolution of the virus with new variants and changes in viral behavior; the public health policies with  implementation, adjustment and measures continuously influence disease dynamics; the dynamics of immunity, including the duration of vaccine-induced immunity and natural immunity, are evolving, affecting disease spread and transition rates. Therefore, the need of non-stationary transition probabilities for our study.
\subsection{Non-stationary Transition Probabilities}\label{SectNonStation}
In contrast to stationary transition probabilities,	non-stationary transition probabilities change over time. This means that the likelihood of moving from one state to another can vary due to factors such as time, evolving disease dynamics, intervention strategies, and seasonal effects. In this study, we focus on time as the primary factor influencing our non-stationary transition probabilities. Specifically, we model the various probabilistic parameters as functions of time to achieve a more accurate and realistic representation of disease dynamics and the impact of interventions. In what follows, $$\beta(t) = 1 - \tau_{is,r}(t) - \tau_{is,h}(t) - \tau_{is,d}(t).$$

	\begin{enumerate}[(a)]
		\item Transition Probability From State $S$ to State $E$\\
		The probability of landing in State $E$ after leaving State $S$ is given by:
		
		\begin{equation}
		    \begin{array}{rllll}
			\P(E|S,t) &=& \begin{pmatrix}
				|S|_k \\~\\ \delta_1
			\end{pmatrix}
			\left(\dis\frac{(\zeta_{ia,s}(t)+\zeta_{is,s}(t)+\zeta_{h,s}(t))(|E|_k + |I_a|_k+|I_s|_k+|H|_k)}{N-|D|_k}\right)^{\delta_1}\\~\\
			&&\left(1-\dis\frac{(\zeta_{ia,s}(t)+\zeta_{is,s}(t)+\zeta_{h,s}(t))(|E|_k + |I_a|_k+|I_s|_k+|H|_k)}{N-|D|_k}\right)^{|S|_k-\delta_1},
		\end{array}
		\end{equation}
		where
        \begin{equation}\label{nonstat1}
		  \zeta_{ia,s}(t) = \zeta_{ia,s}^0 + a_1te^{-kt},
		\end{equation}
		and $\zeta_{is,s}(t), ~\zeta_{h,s}(t)$ are defined as in \eqref{nonstat1}, with $\zeta_{ia,s}^0$ being the initial contact rate between a susceptible and an asymptomatic individual, $a$ the constant indicating how high $\zeta_{ia,s}$ rises, and $k$ the rate of decrease.
        The function defined in \eqref{nonstat1} was borrowed from Ethan Kigundu's research work, former Summer Academy of Actuarial and Mathematical Sciences (SAAMS) scholar from the 2024 cohort at Morgan State University.
		
		\item Transition Probability From State $E$ to State $I_a$\\
		The probability of landing in State $I_a$ after leaving State $E$ is given by:
		
		\begin{equation}
		    \P(I_a|E,t) = \begin{pmatrix}
			|E|_k \\~\\ \delta_2
		\end{pmatrix}
		\tau_{e,ia}^{\delta_2}(t)(1-\tau_{e,ia}(t))^{|E|_k-\delta_2},
		\end{equation}
		where $\tau_{e,ia}(t)$ is defined as in \eqref{nonstat1}.
		
		\item Transition Probability From State $I_a$ to State $I_s$\\
		The probability of landing in State $I_s$ after leaving State $I_a$ is given by:
		
		\begin{equation}
		    \P(I_s \mid I_a,t) = 
		\frac{
			\displaystyle\sum_{\delta_4 = 0}^{|I_a|_k - \delta_3} 
			\binom{|I_a|_k}{\delta_3, \delta_4, Q_1} 
			\tau_{ia,is}(t)^{\delta_3} 
			\tau_{ia,r}(t)^{\delta_4} 
			(1 - \tau_{ia,is}(t) - \tau_{ia,r}(t))^{Q_1}
		}{
			\dis\sum_{\delta_3 = 0}^{|I_a|_k}
			\sum_{\delta_4 = 0}^{|I_a|_k - \delta_3}
			\binom{|I_a|_k}{\delta_3, \delta_4, Q_1} 
			\tau_{ia,is}(t)^{\delta_3} 
			\tau_{ia,r}(t)^{\delta_4} 
			(1 - \tau_{ia,is}(t) - \tau_{ia,r}(t))^{Q_1}
		},
		\end{equation}
		where $\tau_{ia,is}(t)$ is defined as in \eqref{nonstat1} and 
		
		\begin{equation}\label{nonstat2}
			\tau_{ia,r}(t) = \dis\frac{\tau_{ia,r}^{max}}{1+e^{-\beta_2(t-t_0)}}
		\end{equation}
		with $t_0$ representing the time at the inflection point, $\beta_2$ the growth rate and $\tau_{ia,r}^{max}$ the maximum possible value of $\tau_{ia,r}$.
		
		\item Transition Probability From State $I_a$ to State $R$\\
		The probability of landing in State $R$ after leaving State $I_a$ is given by:
		
		\begin{equation}
		    \P(R\mid I_a, t) = 
	\frac{
		\displaystyle\sum_{\delta_3 = 0}^{|I_a|_k - \delta_4} 
		\binom{|I_a|_k}{\delta_3, \delta_4, Q_1} 
		\tau_{ia,is}(t)^{\delta_3} 
		\tau_{ia,r}(t)^{\delta_4} 
		(1 - \tau_{ia,is}(t) - \tau_{ia,r}(t))^{Q_1}
	}{
		\dis\sum_{\delta_4 = 0}^{|I_a|_k}
		\sum_{\delta_3 = 0}^{|I_a|_k - \delta_4}
		\binom{|I_a|_k}{\delta_3, \delta_4, Q_1} 
		\tau_{ia,is}(t)^{\delta_3} 
		\tau_{ia,r}(t)^{\delta_4} 
		(1 - \tau_{ia,is}(t) - \tau_{ia,r}(t))^{Q_1}
	},
		\end{equation}
		where $\tau_{ia,is}(t)$ and $\tau_{ia,r}(t)$ are defined as in \eqref{nonstat1} and \eqref{nonstat2} respectively.
		
		\item Transition Probability From State $I_s$ to State $R$\\
		The probability of landing in State $R$ after leaving State $I_s$ is given by:
		
        \begin{equation}
			\P (R|I_s,t) = \frac{
		\displaystyle\sum_{\delta_6 = 0}^{|I_s|_k - \delta_5}\sum_{\delta_7 = 0}^{|I_s|_k - \delta_5-\delta_6} 
		\binom{|I_s|_k}{\delta_5, \delta_6, \delta_7, Q_2} 
		\tau_{is,r}(t)^{\delta_5} 
		\tau_{is,h}(t)^{\delta_6}\tau_{is,d}(t)^{\delta_7} 
		\beta(t)^{Q_2}
	}{
		\dis\sum_{\delta_5 = 0}^{|I_s|_k}
		\sum_{\delta_6 = 0}^{|I_s|_k - \delta_5}\sum_{\delta_7 = 0}^{|I_s|_k - \delta_5 - \delta_6}
		\binom{|I_s|_k}{\delta_5, \delta_6, \delta_7, Q_2} 
		\tau_{is,r}(t)^{\delta_5} 
		\tau_{is,h}(t)^{\delta_6}\tau_{is,d}(t)^{\delta_7} 
		\beta(t)^{Q_2}
	}
        \end{equation}

		where $\tau_{is,r}(t)$ is defined as in \eqref{nonstat2} and $\tau_{is,h}(t), ~ \tau_{is,d}(t)$ are defined as in \eqref{nonstat1}.
		
		\item Transition Probability From State $I_s$ to State $H$\\
		The probability of landing in State $H$ after leaving State $I_s$ is given by:
		
		\begin{equation}
			\P (H|I_s,t) = \frac{
		\displaystyle\sum_{\delta_5 = 0}^{|I_s|_k - \delta_6}\sum_{\delta_7 = 0}^{|I_s|_k - \delta_5-\delta_6} 
		\binom{|I_s|_k}{\delta_5, \delta_6, \delta_7, Q_2} 
		\tau_{is,r}(t)^{\delta_5} \tau_{is,h}(t)^{\delta_6}\tau_{is,d}(t)^{\delta_7} 
		\beta(t)^{Q_2}
	}{
		\dis\sum_{\delta_6 = 0}^{|I_s|_k}
		\sum_{\delta_5 = 0}^{|I_s|_k - \delta_6}\sum_{\delta_7 = 0}^{|I_s|_k - \delta_5 - \delta_6}
		\binom{|I_s|_k}{\delta_5, \delta_6, \delta_7, Q_2} 
		\tau_{is,r}(t)^{\delta_5} \tau_{is,h}(t)^{\delta_6}\tau_{is,d}(t)^{\delta_7} 
		\beta(t)^{Q_2}
	}
		\end{equation}
		where $\tau_{is,r}(t)$ is defined as in \eqref{nonstat2} and $\tau_{is,h}(t), ~ \tau_{is,d}(t)$ are defined as in \eqref{nonstat1}.
		
		\item Transition Probability From State $I_s$ to State $D$\\
		The probability of landing in State $D$ after leaving State $I_s$ is given by:
		
		\begin{equation}
			\P (D|I_s,t) = \frac{
		\displaystyle\sum_{\delta_5 = 0}^{|I_s|_k - \delta_7}\sum_{\delta_6 = 0}^{|I_s|_k - \delta_5-\delta_7} 
		\binom{|I_s|_k}{\delta_5, \delta_6, \delta_7, Q_2} 
		\tau_{is,r}(t)^{\delta_5} \tau_{is,h}(t)^{\delta_6}\tau_{is,d}(t)^{\delta_7} 
		\beta(t)^{Q_2}
	}{
		\dis\sum_{\delta_7 = 0}^{|I_s|_k}
		\sum_{\delta_5 = 0}^{|I_s|_k - \delta_7}\sum_{\delta_6 = 0}^{|I_s|_k - \delta_5 - \delta_7}
		\binom{|I_s|_k}{\delta_5, \delta_6, \delta_7, Q_2} 
		\tau_{is,r}(t)^{\delta_5} \tau_{is,h}(t)^{\delta_6}\tau_{is,d}(t)^{\delta_7} 
		\beta(t)^{Q_2}
	}
		\end{equation}
		where $\tau_{is,r}(t)$ is defined as in \eqref{nonstat2} and $\tau_{is,h}(t), ~ \tau_{is,d}(t)$ are defined as in \eqref{nonstat1}.
		
		\item Transition Probability From State $H$ to State $D$\\
		The probability of landing in State $D$ after leaving State $H$ is given by:
		
		\begin{equation}
		    \P (D|H,t) = \frac{
		\displaystyle\sum_{\delta_9 = 0}^{|H|_k - \delta_8} 
		\binom{|H|_k}{\delta_8, \delta_9, |H|_k - \delta_8 - \delta_9} 
		\tau_{h,r}(t)^{\delta_9} 
		\tau_{h,d}(t)^{\delta_8} 
		(1 - \tau_{h,r}(t) - \tau_{h,d}(t))^{|H|_k - \delta_8 - \delta_9}
	}{
		\dis\sum_{\delta_8 = 0}^{|H|_k}
		\sum_{\delta_9 = 0}^{|H|_k - \delta_8}
		\binom{|H|_k}{\delta_8, \delta_9, |H|_k - \delta_8 - \delta_9} 
		\tau_{h,r}(t)^{\delta_9} 
		\tau_{h,d}(t)^{\delta_8} 
		(1 - \tau_{h,r}(t) - \tau_{h,d}(t))^{|H|_k - \delta_8 - \delta_9}
	}
		\end{equation}
		where $\tau_{h,d}(t)$ is defined as in \eqref{nonstat1} and 
        \begin{equation}\label{nonstat3}
			\tau_{h,r}(t) = \tau_{h,r}^0 - a_2te^{-kt},
		\end{equation}
		with $\tau_{h,r}^0$ being the initial probability of recovering after being discharged from ICU, $a$ the constant indicating how high $\tau_{h,r}$ rises, $k$ the rate of decrease. 
		
		\item Transition Probability From State $H$ to State $R$\\
		The probability of landing in State $R$ after leaving State $H$ is given by:
		
		\begin{equation}
		    \P (R|H,t) = \frac{
		\displaystyle\sum_{\delta_8 = 0}^{|H|_k - \delta_9} 
		\binom{|H|_k}{\delta_8, \delta_9, |H|_k - \delta_8 - \delta_9} 
		\tau_{h,r}(t)^{\delta_9} 
		\tau_{h,d}(t)^{\delta_8} 
		(1 - \tau_{h,r}(t) - \tau_{h,d}(t))^{|H|_k - \delta_8 - \delta_9}
	}{
		\dis\sum_{\delta_9 = 0}^{|H|_k}
		\sum_{\delta_8 = 0}^{|H|_k - \delta_9}
		\binom{|H|_k}{\delta_8, \delta_9, |H|_k - \delta_8 - \delta_9} 
		\tau_{h,r}(t)^{\delta_9} 
		\tau_{h,d}(t)^{\delta_8} 
		(1 - \tau_{h,r}(t) - \tau_{h,d}(t))^{|H|_k - \delta_8 - \delta_9}
	}
		\end{equation}
		where $\tau_{h,d}(t)$ and $\tau_{h,r}(t)$ are defined by \eqref{nonstat1} and \eqref{nonstat3} respectively.
		
		\item Transition Probability From State $R$ to State $S$\\
		The probability of landing in State $S$ after leaving State $R$ is given by:

        \begin{equation}
            \P(S|R,t) = 1,
        \end{equation}
        since no individual gains immunity after recovering from the disease.
	\end{enumerate}

\section{COVID-19 Dynamics Over 365 Days}\label{example}
Consider a population of 6,500,000 individuals, initially distributed as shown in Table \ref{Init_Cond}. The constants within the probabilistic functions have also been considered in Table \ref{Const_val}, to enable the execution of the dynamic simulation.

\begin{table}[h!]
\centering
  \caption{Initial conditions of the state variables}
	\begin{tabular}{|c|c|}
		\hline
		Initial conditions & Values \\
		\hline
		$S(0)$ & $5654000$ \\
		\hline
		$E(0)$ & $817097$ \\
		\hline
		$I_a(0)$ & $22750$ \\
		\hline
		$I_s(0)$ & $4550$ \\
		\hline
		$H(0)$ & $617$ \\
		\hline
		$R(0)$ & $357$ \\
		\hline
		$D(0)$ & $629$ \\
		\hline
	\end{tabular}\label{Init_Cond}
    \vspace{.5cm}
 \centering
 \caption{Constants within the probabilistic functions}
	\begin{tabular}{|c|c|}
		\hline
		constants & Values \\
        \hline
        $t_0$ & $0$ \\
		\hline
        $\tau_{r,s}(t)$ & $1$ \\
        \hline
		$\beta_2$	& $0.5$ \\
		\hline
		$k$	& $5$ \\
		\hline
  	$a_1$	& $1$ \\
		\hline
		$a_2$	& $2$ \\
		\hline
		$\zeta_{ia,s}^{0}$	& $0.15$ \\
		\hline
  	$\tau_{h,r}^{0}$	& $0.15$ \\
		\hline
        $\tau_{ia,r}^{max}$ & $0.04$ \\
        \hline
        $\tau_{is,r}^{max}$ & $0.08$ \\
        \hline
	\end{tabular}\label{Const_val}
 \end{table}

\noindent
The dynamic simulation below, over a 365-day period illustrates how the varying parameter $t$ influences the progression of the disease, with a particular focus on the susceptible and deceased populations and their impact on intermediate states.

\begin{itemize}
    \item \textbf{Observation 1 (Figure \ref{Low_t}):}
    \begin{figure}[h!]
	\centering
	\includegraphics[height=8.5cm]{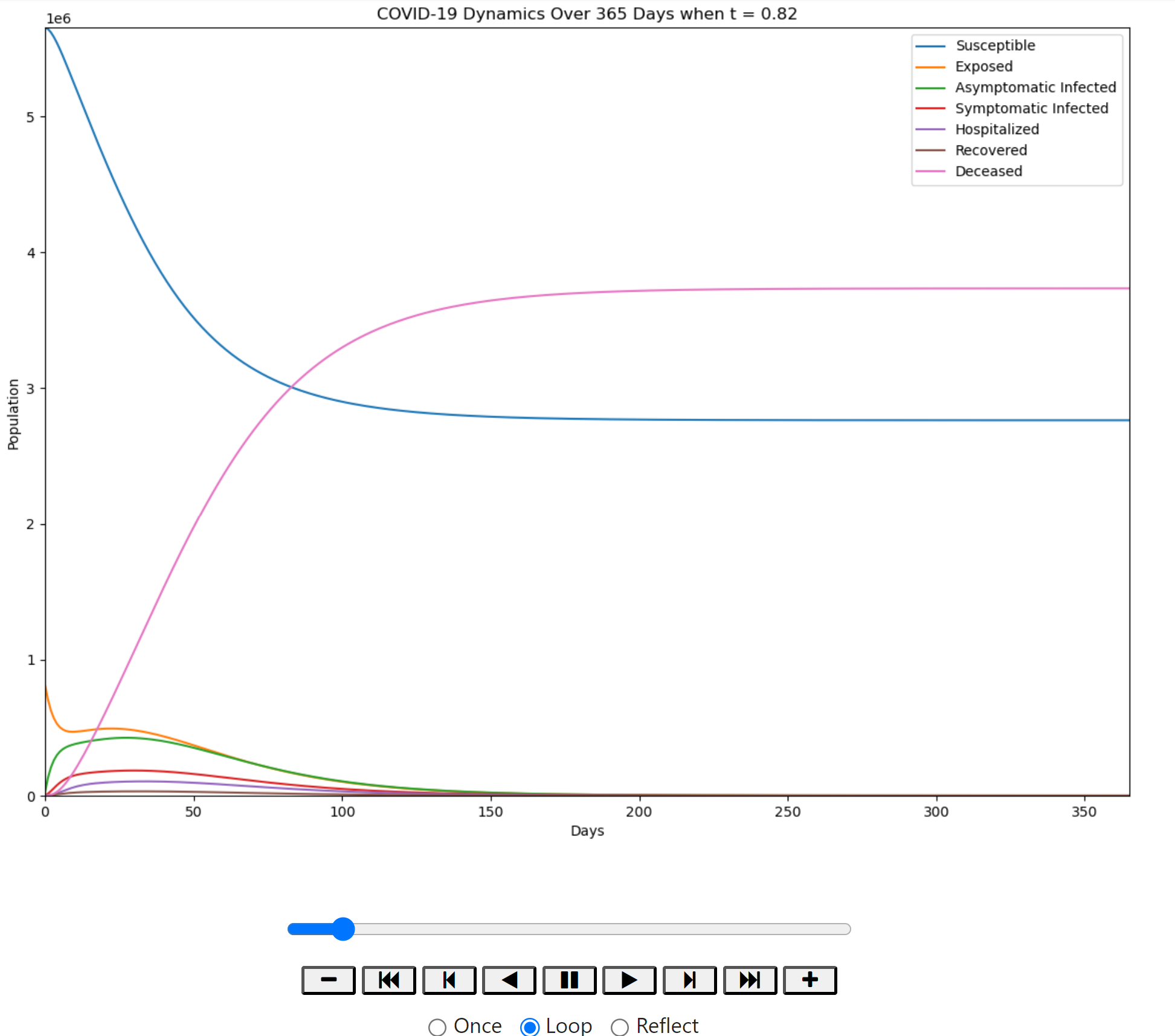}
	\caption{Low Values of $t$ (from $0$ up to $2.18$)}
	\label{Low_t}
    \end{figure}
With low values of $t$, the severity of the disease, combined with a slow recovery rate, leads to a sharp increase in the deceased population and a rapid decrease in the susceptible population. Initially, the number of susceptibles remains higher than the number of deceased, but as the disease progresses, the two populations overlap, and the number of deceased eventually surpasses that of the susceptibles, indicating high lethality and a prolonged impact of the disease. This rapid decline in susceptibles and sharp rise in deceased create a cascading effect on intermediate states, such as the infected and recovered populations. The reduction in susceptibles decreases the pool of individuals available for new infections, while the increase in deceased reduces the number of individuals who can transition to other compartments.

    \item \textbf{Observation 2 (Figure \ref{High_t}):}
    \begin{figure}[h!]
	\centering
	\includegraphics[height=8.5cm]{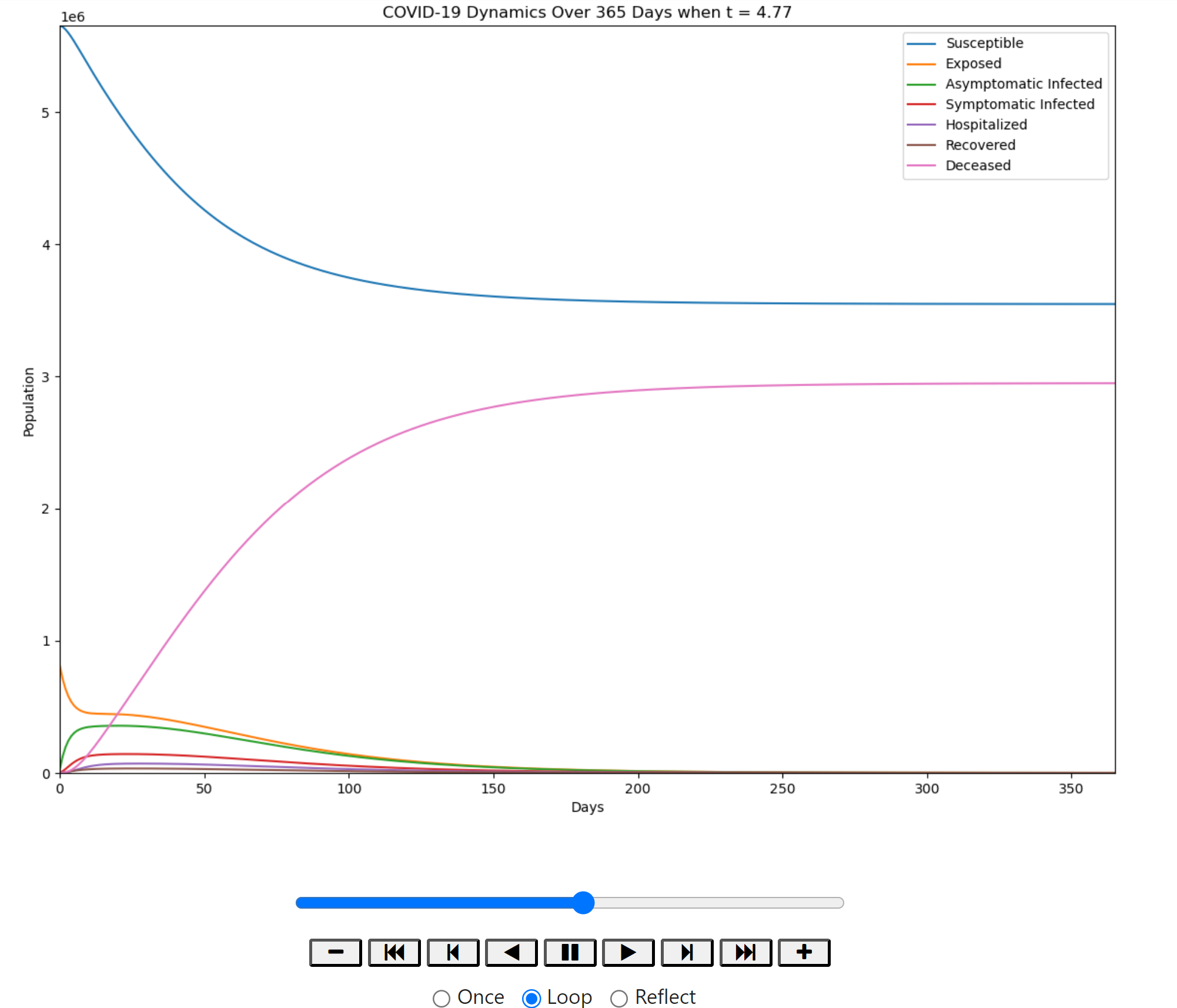}
	\caption{Moderate to High Values of $t$ ($2.18$ and above)}
	\label{High_t}
    \end{figure}
When $t$ reaches $2.18$ and beyond, the dynamics settle. The susceptible population remains consistently larger than the deceased population, due to a slower increase and decrease in the deceased and susceptible compartments, respectively. This phase signifies a balance between disease spread and recovery, resulting in a more controlled epidemic progression. By approximately day 250, the system reaches a steady state, with variations in all compartments approaching zero. Both the susceptible and deceased populations balance, indicating that the epidemic has subsided and the populations are no longer undergoing significant changes.
\end{itemize}

\noindent
These plots illustrate the progression of individuals across different compartments of a COVID-19 model over time (in days), offering valuable insights for understanding and predicting disease dynamics, and for planning public health interventions. Overall, as the probabilistic function increases (when $t$ is in the range $[0, 0.4)$), there is a rapid decline in the susceptible population, reflecting a high rate of new infections, while the deceased population rises quickly, indicating a severe impact of the disease and a high transition rate from infection to death. This rapid decline in susceptibles and fast increase in deceased population reduce the number of active infections and limit the size of the recovered population, thereby influencing the overall dynamics. When the probabilistic function decreases (for $t$ in the range $[0.4, 4.2)$), the decline in the susceptible population slows, indicating a reduction in the infection rate, and the increase in the deceased population also slows, reflecting a less aggressive disease progression. This leads to a more balanced dynamic, with intermediate compartments like the infected population maintaining a more consistent size. Finally, when the probabilistic function becomes constant (approaching $0.15$ for $t$ in the range $[4.2, \infty)$), the sizes of the susceptible and deceased populations settle, indicating that the system has reached an equilibrium with no significant new infections or deaths occurring. This constancy suggests that other compartments, such as the infected and recovered populations, have also leveled off, leading to a long-term steady state in the epidemic's dynamics.

\section{Concluding remarks}\label{conclusion}
This study develops transition probability models using the chain-binomial approach, tailored for MDPs with non-stationary transition probabilities. Incorporating non-stationary transition probabilities offers significant advantages in accurately capturing the evolving dynamics of COVID-19. Unlike stationary models, which assume constant transition probabilities, non-stationary models account for changes over time due to factors such as evolving virus variants, seasonal variations, and the impact of public health interventions. By reflecting these time-dependent changes, non-stationary transition probabilities provide a more nuanced understanding of disease progression and help policymakers adapt strategies effectively as conditions evolve.\\

\noindent Our findings emphasize that while stationary models provide a stable baseline for evaluating long-term interventions and resource allocation, they may be inadequate in scenarios where disease patterns fluctuate over time. In contrast, the non-stationary models, built upon the chain-binomial framework, offer the adaptability necessary for real-time forecasting and timely policy adjustments. This approach enables more accurate modeling of the pandemic's dynamics and improves the efficacy of response strategies. This ensures that policies remain effective as circumstances shift, optimizing resource allocation and potentially leading to cost savings by avoiding both over- and under-preparation scenarios. Therefore, integrating non-stationary transition probabilities into MDP frameworks enhances the ability to implement timely and effective interventions, improving overall public health outcomes.\\

\noindent
In conclusion, both stationary and non-stationary transition probabilities play distinct yet complementary roles in optimizing decision-making for managing the dynamics of COVID-19. Stationary models provide simplicity and consistency, making them well-suited for long-term planning and resource allocation. They offer a stable reference point, which is valuable for assessing overall trends and guiding strategic decisions. Conversely, non-stationary models introduce the flexibility and adaptability necessary for dynamic response and accurate forecasting. By accounting for time-varying factors, these models allow for real-time adjustments based on evolving data, which is crucial for timely and effective policy-making. Integrating both stationary and non-stationary approaches creates a balanced framework that leverages the stability of stationary models alongside the adaptability of non-stationary models. This integrated approach enhances public health outcomes by providing a comprehensive tool for both strategic planning and responsive action, improving economic efficiency in managing the COVID-19 pandemic. Future research focusing on cost-benefit analysis will further refine these models, enhancing their utility and effectiveness in crafting optimized pandemic responses.

\section*{Conflict of interest statement}
The authors declare that there is no conflict of interest regarding the publication of this article.

\section*{Availability of data and material }
The datasets used and/or analysed during the current study are available from the corresponding author on request.

\section*{Funding}
This work was funded by the U.S. Department of Energy (DOE) under grant DE-AC02-05CH11231 through the Biopreparedness Research Virtual Environment (BRaVE) program,  "EMERGE: ExaEpi for Elucidating Multiscale Ecosystem Complexities for Robust, Generalized Epidemiology."

\newpage
\bibliographystyle{amsplain}

\end{document}